\documentclass[aps,pra,amsmath,twocolumn]{revtex4}

\usepackage{amsmath,amssymb}
\usepackage{graphicx}
\usepackage{color}
\usepackage{xspace}
\usepackage{mathtools}

\newcommand{\bra}[1]{\langle #1|}
\newcommand{\ket}[1]{|#1\rangle}

\newcommand{\e}{\mathrm{e}}

\newcommand{\der}[2]{\frac{d #1}{d #2}}
\newcommand{\dder}[2]{\frac{d^2 #1}{d #2^2}}

\newcommand{\QAO}{\textsc{qao}\xspace}

\newcommand{\gmin}{g_\mathrm{min}}

\renewcommand{\epsilon}{\varepsilon}

\preprint{Preprint}

\begin{document}

\title{Discrepancies between Asymptotic and Exact Spectral Gap Analyses 
of Quantum Adiabatic Barrier Tunneling}

\author{Lucas~T.~Brady}
\affiliation{{Department of Physics, University of California, Santa Barbara, CA 93106-5110, USA}}
\author{Wim~van~Dam}
\affiliation{{Department of Computer Science, Department of Physics, University of California, Santa Barbara, CA 93106-5110, USA}}

\date{\today}
\begin{abstract}
We study the asymptotic behavior of the spectral gap of simple barrier tunneling problems, 
which are related to using quantum annealing to find the global optimum of cost functions
defined over $n$ bits. 
Specifically we look at the problem of having an  $n$ qubit system tunnel through a barrier of width and height
proportional to $n^\alpha$. 
We show that with these quantum annealing problems, the asymptotic, $n\to\infty$, 
behavior of the spectral gap does not accurately describe the
behavior of the gap at finite $n$ until extremely large values of $n$
($n>10^{12}$). We prove that this deficiency of the asymptotic expression is a
feature of simple one-dimensional tunneling problems themselves, casting doubt
on whether asymptotic analysis is an appropriate tool for studying tunneling
problems in quantum annealing for reasonably sized systems.
\end{abstract}

\maketitle

\section{Introduction}
\label{sec:intro}
Quantum tunneling is one of the major aspects of quantum mechanics that sets it
apart from classical mechanics. The ability of a quantum system to enter a
classically disallowed region and, through it, reach an energetically favorable
region has inspired many applications. In this article, we are concerned
with tunneling as it pertains to quantum annealing, which is a quantum algorithm
that uses quantum effects, such as tunneling, to relax a system into a desired
ground state.

Quantum adiabatic optimization (\QAO) \cite{Farhi2000} is a form of quantum annealing that solves optimization problems by adiabatically changing the Hamiltonian for a system from one with a known ground state to one with the optimization solution encoded in its ground state.  Many studies \cite{Farhi2002, Farhi2008, Martonak, Hastings, Heim, Boxio2, Battaglia, Crosson, Harrow, Muthukrishnan, Brady, Kong} have sought to characterize what, if any, speed-up this algorithm has over classical alternatives such as simulated annealing or simulated quantum annealing.  A popular belief among the \QAO community is that at least part of quantum annealing's power comes from its ability to tunnel through barriers more easily than classical algorithms can go over them, but several recent results \cite{Crosson, Harrow, Jiang, Kong, Brady} have shown that tunneling alone is not enough to ensure a quantum speed-up.

Nevertheless, generalized \QAO is universal for quantum computation \cite{Aharonov}, and an understanding of tunneling in a quantum annealing setting can shed light on the power and limitations of \QAO, even if tunneling alone is not enough for a quantum speed-up.  To that end, this article studies quantum tunneling as part of \QAO in the simplest possible situation: tunneling in position-space through a one-dimensional barrier.  In a previous article \cite{Brady2}, we showed that this simple problem is equivalent to barrier tunneling of $n$ qubits in a bit symmetric setting in the large $n$ limit.

Despite the simplicity of one-dimensional position-space tunneling, we find that an asymptotic analysis of \QAO spectral gaps for $n\to\infty$, which give bounds on the adiabatic run time, fail to accurately describe the behavior of the tunneling for moderately large $n$.  We find that $n> 10^{12}$ is needed to ensure that asymptotic expressions accurately describe behavior, to the point where polynomial and exponential run time scaling can be confused for lower $n$.  Given the current and near-term sizes of quantum computer implementations \cite{Boxio3}, this failure of asymptotics casts doubt on the efficacy of using asymptotics to analyze the output of these systems or to predict their capabilities.

In Section~\ref{sec:background} we provide a brief review of \QAO and introduce the details of the barrier tunneling problem to be studied.  We mention both the original hypercube problem and the simpler continuous one-dimensional problem.

Section~\ref{sec:trough} gives the first evidence that the asymptotic limit may not be accurate even for moderately large $n$.  We show that the failure of the asymptotics is the same in both the hypercube and one-dimensional cases, proving that the failure is in the tunneling problem itself and not in the approximation scheme from \cite{Brady2}.

We compare the exact spectral gap to its asymptotic approximation in Section~\ref{sec:exact_scaling} to see where the issues discussed in the previous section arise.  This section also provides estimates for the $n$ necessary to see asymptotic behavior and explores what effect next-order corrections to the asymptotics have on the analysis.

Section~\ref{sec:scaling} uses properties of the derivatives of log-log data to analyze how well a curve approximates a power law versus and exponential fit.  Furthermore, we show how easily finite data for tunneling could lead to a mislabeling of the scaling behavior of the algorithm.  Finally, Section~\ref{sec:conc} concludes and provides an outlook for the future of such asymptotic analysis.

\section{Tunneling in Quantum Adiabatic Optimization}
\label{sec:background}

Quantum annealing derives its name by comparison with classical simulated annealing, which starts a system at a high temperature and then slowly lowers the temperature, trying to relax the system into its ground state.  Quantum annealing works on the same principle except it uses quantum effects instead of temperature.

\subsection{Quantum Adiabatic Optimization}

As it is usually formulated \cite{Farhi2000}, the quantum annealing Hamiltonian is a linear time interpolation between two static Hamiltonians:
\begin{equation}
      \hat{H}(s) = (1-s)\hat{H}_0+s\hat{H}_1,
\end{equation}
where $s=t/\tau$ is the time parameter, $t$, divided by the total predetermined runtime of the algorithm, $\tau$.  $\hat{H}_0$ is the initial Hamiltonian and is usually chosen to have an easily prepared ground state.  $\hat{H}_1$ encodes the solution to an optimization problem in its ground state configuration.  The initial formulation of \QAO \cite{Farhi2000} set this in a Hilbert space of $n$ qubits where $\hat{H}_0$ is a uniform field in the negative $x$ direction:
\begin{equation}
      \hat{H}_0 = -\sum_{i=1}^n \sigma_x^{(i)},
\end{equation}
and $\hat{H}_1$ is diagonal the in $z$ basis with a cost function encoded along its diagonal:
\begin{equation}
      \hat{H}_1 = \sum_{z\in\{0,1\}^n} f(z)\ket{z}\bra{z}.
\end{equation}

These canonical choices for $\hat{H}_0$ and $\hat{H}_1$ make the the algorithm easier to create and analyze, but they also ensure that $\hat{H}(s)$ is stoquastic, meaning its off-diagonal terms are all non-positive.  The universality proof \cite{Aharonov} for \QAO requires non-stoquastic Hamiltonians, so it is currently unclear how much power stoquastic Hamiltonians have in \QAO.

Key to \QAO is that the algorithm must run adiabatically, where adiabatically means that the total time, $\tau$, scales according to a combination of matrix norms of time derivatives of $\hat{H}(s)$ and the spectral gap, $g(s)$, between the ground state and first excited state.  Typically, the adiabatic condition is stated as adiabaticity is ensure if $\tau\gg \max_{s\in[0,1]}||\der{\hat{H}(s)}{s}||/(g(s))^2$.  The derivation of this version of the adiabatic theorem is subtly incorrect, which has led to more robust statements and derivations of an adiabatic condition \cite{Jansen}, but all of these more accurate conditions claim that the worst-case bound on $\tau$ depends on matrix norms, which in our case are linear in $n$, and the inverse of $\gmin \coloneqq \min_{s\in[0,1]} g(s)$.  Therefore, the majority of this paper focuses on the calculation of $\gmin$ since this is the primary changing quantity that determines the worst-case bound on the run time of \QAO.

\subsection{Symmetric Barrier Tunneling Problem}
\label{sub:problem}

Our barrier tunneling problem is inspired by tunneling using $n$ qubits in a bit-symmetric setting.  A thin spike version of this bit-symmetric problem was originally proposed by Farhi et al. \cite{Farhi2002}, and the general bit-symmetric barrier tunneling problem has been analyzed extensively by other groups \cite{Reichardt,Crosson,Brady,Harrow,Kong,Brady2,Jiang,Goldstone}.  The final cost function $f(z)$ is a function only of the Hamming weight $|z|$ as $f(z) = |z| + b(|z|)$ where $b(w)$ is a function centered around $w=n/4$ with width and height proportional to $n^\alpha$.

In a previous article \cite{Brady2}, we showed that this Hamiltonian can be mapped onto the symmetric $(n+1)$-dimensional subspace where it describes a single spin $1/\epsilon\coloneqq n/2$ particle.  Furthermore, using a modified version of the Villain transformation \cite{Villain}, we showed that this system can be mapped, in the $n\to\infty$ limit, onto a semiclassical wave equation in one continuous variable $x$.  This mapping only works for low-lying energy states, but because we only care about the spectral gap between the ground state and first excited state, this restriction is acceptable.  In \cite{Brady2}, we restrict ourselves to a square barrier so that our final, semiclassical Schr\"odinger equation becomes
\begin{equation}
      \label{eq:main_DE}
      \dder{\psi}{x} = \epsilon^{-2}\left[V(x)-\epsilon c E\right]\psi(x),
\end{equation}
where
\begin{equation}
      V(x) = \begin{cases}
            \omega\epsilon^{1-\alpha}&\text{if}~-a<x<a\\
            \omega^2 x^2 & \text{otherwise}
      \end{cases}
\end{equation}
The wave function $\psi(x)$ is the continuous version of the eigenvector components in the spin-$n/2$ problem, and $E$ is the eigenenergy.  The variable $x$ is a rescaled and shifted version of the Hamming weight, taken in the continuum limit, and the barrier has also been rescaled to have width $\epsilon^{1-\alpha} \eqqcolon 2a$ and height $\omega\epsilon^{1-\alpha}$.  The constants can either be taken as arbitrary or set to $c=8/(3(\sqrt{3}-1))$ and $\omega=4/3$ if we want a direct correlation to the qubit problem.

While Eq.~\ref{eq:main_DE} was derived from the $n$ qubit problem, it can also stand on its own, describing the critical tunneling moment in a one-dimensional continuous system.  In this context, we just have a particle moving in a one-dimensional quadratic well with a barrier in the middle.  For most of this article, we treat the Schr\"odinger equation in Eq.~\ref{eq:main_DE} as its own independent problem.

The continuous well differential equation, Eq.~\ref{eq:main_DE}, can be solved using exponentials and parabolic cylinder functions, $D_\nu(y)$, up to a transcendental equation for the eigenenergies:
\begin{align}
      \label{eq:trans_con}
      &k_\pm D_{\nu_\pm}\left(\sqrt{\frac{2\omega}{\epsilon}}a\right)\left(\e^{k_\pm a}\mp \e^{-k_\pm a}\right)=\nonumber\\
            &\sqrt{\frac{2\omega}{\epsilon}} D'_{\nu_\pm}\left(\sqrt{\frac{2\omega}{\epsilon}}a\right)\left(\e^{k_\pm a}\pm \e^{-k_\pm a}\right),
\end{align}
where $\nu_\pm\coloneqq \frac{cE_{\pm}}{2\omega}-\frac{1}{2}$, $k_{\pm}\coloneqq \sqrt{\omega\epsilon^{-1-\alpha}-\epsilon^{-1} cE_{\pm}}$, and $E_{\pm}$ refers to even ($+$) or odd ($-$) energy solutions.  In the limit of $n\to\infty$, Eq.~\ref{eq:trans_con} can be solved for the lowest two energy levels, resulting in a gap expression
\begin{equation}
      \label{eq:gap_poly}
      \gmin = \frac{8\omega^{1/2}}{c\sqrt{\pi}}\epsilon^{2\alpha-1/2}
\end{equation}
for $1/4<\alpha<1/3$, and 
\begin{equation}
      \label{eq:gap_exp}
      \gmin = \frac{16 \omega}{c\sqrt{\pi}} \epsilon^{\alpha/2} \exp\left(-\sqrt{\omega}\epsilon^{\frac{1}{2}-\frac{3}{2}\alpha}\right)
\end{equation}
for $1/3<\alpha<1/2$.  These asymptotic gap expressions match results for the original symmetric hypercube tunneling problem from other groups using various methods \cite{Farhi2002,Goldstone,Kong,Jiang}.

\subsection{Definitions of Gap Analyses}

In Subsection~\ref{sub:problem} we outline three steps in the approximation procedure from which the gap can be extracted.  In the rest of the paper we will refer back to these three different models, so we will take a moment here to label these three models and define exactly what we mean by each term.

\begin{itemize}

\item {\bf Discrete analysis}: The original $n$ qubit hypercube problem consists of a discrete Hilbert space that can be interpreted in terms of Hamming weight or spin states.  The gap can be extract from this problem through exact diagonalization of the $(n+1)$-dimensional symmeterized problem.  While the discrete problem was the original inspiration for this work, much of the subsequent text will focus exclusively on the continuous and asymptotic gaps.

\item {\bf Continuous analysis}: After the Villain transformation, the discrete problem transforms into a parabolic well potential for a continuous, position-like variable.  The Schr\"odinger equation governing this continuous problem is shown in Eq.~\ref{eq:main_DE}, and the exact spectral gap can be extracted by numerically solving the transcendental equation for the energies, Eq.~\ref{eq:trans_con}.  For most of this article, the continuous problem will be treated as an independent problem, and most of our efforts will focus on comparing its exact spectral gap to the asymptotic limit.

\item {\bf Asymptotic analysis}: In the large $n$ limit, the asymptotic expression for the spectral gap of the discrete and continuous problems can be derived.  We did this derivation in a previous paper \cite{Brady2}, and the gap expressions are reproduced in Eqs.~\ref{eq:gap_poly} and \ref{eq:gap_exp}.  This version of the gap is the true asymptotic limit of the discrete and continuous gaps, but much of this article will be focused on how large $n$ must be for the asymptotic limit to be a good approximation of the true gap.

\end{itemize}

\section{Scaling Power Comparison}
\label{sec:trough}

When analyzing an algorithm numerically or experimentally, a typical question to ask is how its run time, or in our case the gap, scales with the input size $n$.  Therefore, in this section we analyze the gap numerically mostly ignoring our asymptotic expressions from Eqs.~\ref{eq:gap_poly} and \ref{eq:gap_exp} and try to determine what asymptotic information we would extract from numerical data of $n$ vs. $\gmin$.  We then compare these finite $n$ extrapolations to see how close these results are to our analytic expressions for the asymptotic behavior of $\gmin$.

\begin{figure}
      \includegraphics[width=0.48\textwidth]{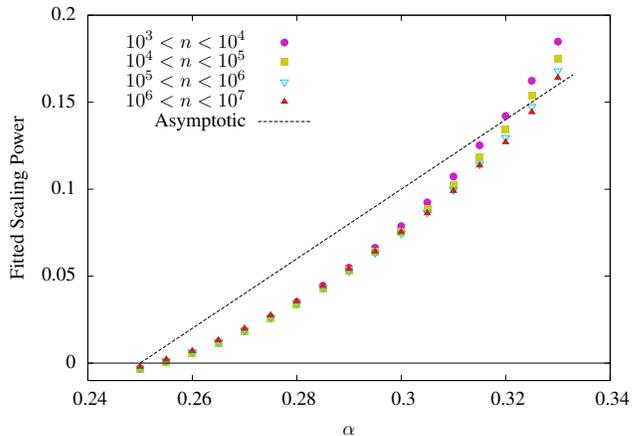}
      \caption{            
                  Fitted scaling exponent of $\gmin$ for the symmetric qubit
                  tunneling problem with a binomial shaped barrier. Each data
                  point shows what scaling power, $p$, is extracted from a fit
                  to $\gmin = A n^{-p}$, assuming the fit is done on data from a
                  finite range of $n$ versus the scaling power of the barrier
                  size, $\alpha$. Each type of point represents data
                  binned from a different range of $n$, and the dashed line
                  represents the power $p$ predicted by the asymptotic
                  expression. Based on this data alone, the scaling of $\gmin$
                  does not seem to approach the asymptotic expression.
      }
      \label{fig:trough_qubit}
\end{figure}

We use symmetry to reduce the discrete qubit Hamiltonian to a tridiagonal $(n+1)$-dimensional matrix and then numerically diagonalize it for a finite range of $n$ before performing power law fits of the form $\gmin = A n^p$ in order to extract the exponent $p$.  Fig.~\ref{fig:trough_qubit} outlines the resulting $p$ values for these fits done on simulations with $1/4<\alpha<1/3$, leading to polynomial scaling asymptotically.

Also, note that the data in Fig~\ref{fig:trough_qubit} comes from a problem with a binomial shaped barrier rather than the normal square barrier we consider.  To collect these data points, we resitricted ourselves to $n$ such that the discrete width of the barrier has just increased, so the binomial barrier's full width, which grows with $n^{2\alpha}$, allows us to collect more data points than the square barrier, whose width grows with $n^{\alpha}$.  Therefore, we used the binomial barrier in order to collect more data points and create a cleaner data set.

In Fig.~\ref{fig:trough_qubit}, we show $p$ values for fits done on different bins of $n$ data, so the red filled triangle should most closely approximate the asymptotic value of $p$.  Unfortunately, the scaling powers do not seem to approach their asymptotic values, shown as a dashed black line in the figure, even for the relatively large problem size of one million qubits.  The lack of agreement with the asymptotics, even for relatively large numbers of qubits casts doubt on the validity of the asymptotic analysis techniques used in \cite{Brady2}.

\begin{figure}
      \includegraphics[width=0.48\textwidth]{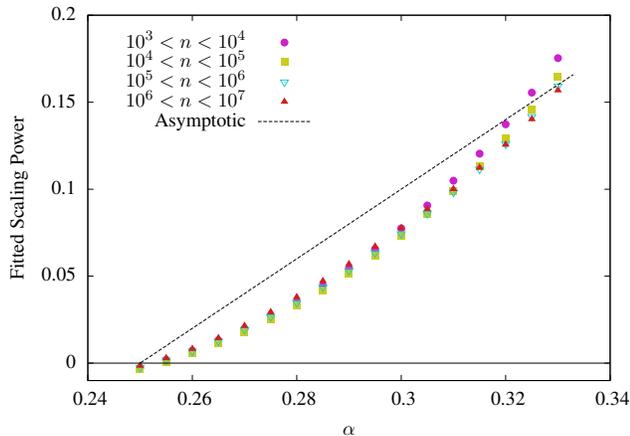}
      \caption{            
                  Fitted scaling exponent of $\gmin$ for the continuous well DE,
                  Eq.~\ref{eq:main_DE}, with a square barrier. Each data
                  point shows what scaling power, $p$, is extracted from a fit
                  to $\gmin = A n^{-p}$, assuming the fit is done on data from a
                  finite range of $n$ versus the scaling power of the barrier
                  size, $\alpha$. Each type of point represents data binned from
                  a different range of $n$, and the dashed line represents the
                  power $p$ predicted by the asymptotic expression. The trough
                  shows the same behavior as seen in
                  Fig.~\ref{fig:trough_qubit}, indicating that the problem in
                  that plot was not in the approximations made for the Villain
                  transformation.
      }
      \label{fig:trough_DE}
\end{figure}

However, the only possible source of error in the asymptotic analysis in \cite{Brady2} is in the Villain transformation and the other assumptions that lead to the derivation of the continuous well DE, Eq.~\ref{eq:main_DE}.  Therefore, if the spectral gap for the differential equation shows the same difficulty in reaching its asymptotic value, we know that extremely lare $n$ are necessary to observe asymptotic behavior because this is a simple tunneling problem, not because the Villain transformation was used in the derivation of these asymptotics.  We do in fact see that if we perform the same procedure that gave us Fig.~\ref{fig:trough_qubit} for numerical $\gmin$ values found from solving Eq.~\ref{eq:trans_con}, we obtain a very similar plot, Fig.~\ref{fig:trough_DE}.

The failure of $\gmin$ to quickly reach its asymptotic scaling even in Fig.~\ref{fig:trough_DE} indicates that there is something special about simple tunneling problems that defies the conventional wisdom that asymptotic behavior sets in quickly.  Based on these figures alone, an experimental or numerical approach to this problem would mislabel the polynomial scaling of this algorithm.  From this point, we focus exclusively on the continuous problem, ignoring the discrete problem that motivated it.  In the next section, we examine what $n$ are necessary to bring the scaling close to the asymptotic expressions.

\section{Comparison of Exact Gap and Asymptotic Expressions}
\label{sec:exact_scaling}

Given the large discrepancy seen in Section~\ref{sec:trough} between finite $n$ runtime scaling and asymptotic scaling, it makes sense to explore what values of $n$ are necessary to get close to the true asymptotic scaling behavior.  The discrepancy seen in the last section is largest in the polynomial scaling region near the middle of the region far from the boundaries with the constant and exponential regions.  Much of the analysis in this section is done for $\alpha=3/10$ which is a simple value in the middle of the polynomial region where this asymptotic discrepancy effect is large.

\subsection{Direct Gap Comparison}

Since our asymptotic analysis of $\gmin$ in Eqs.~\ref{eq:gap_poly} and \ref{eq:gap_exp} is exact enough to get the constant prefactor, we can compare the numerical gap directly to the asymptotic expression to visually and numerically see where agreement occurs.

\begin{figure}
      \includegraphics[width=0.48\textwidth]{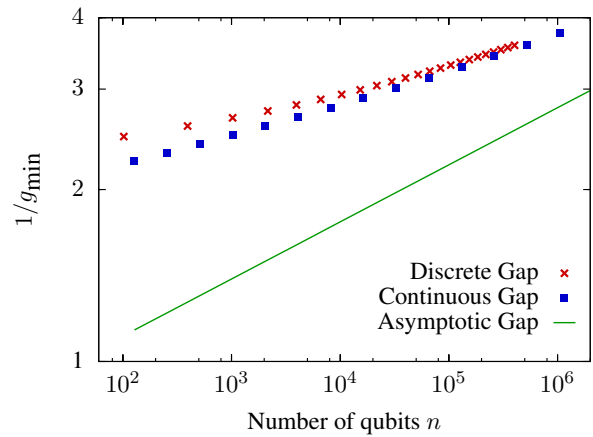}
      \caption{
Spectral gap            $\gmin$ vs.\ $n$ for $\alpha=3/10$ for our three different gap extraction methods with a square barrier.  The discrete gap and continuous gap well approximate each other at these relatively modes $n$, but neither of them are well approximated by the asymptotic expression for the gap.  This difference seems to confirm the failure of the asymptotics observed in Figs.~\ref{fig:trough_qubit} and \ref{fig:trough_DE}.
      }
      \label{fig:comparison}
\end{figure}

In Fig.~\ref{fig:comparison} we show the discrete, continuous, and asymptotic gaps for $\alpha=3/10$ up to $n=10^6$.  Since this is a log-log plot, a straight line corresponds to a power law with the slope equal to the scaling power.  The continuous and discrete gaps approach each other for the reasonable values of $n$ shown in this plot and have very similar slopes.  However, neither the continuous nor the discrete gaps are close to the asymptotic expression, nor do their slopes appear similar to the asymptotic slope.  Fig.~\ref{fig:comparison} shows how far away the actual gap is from the asymptotic expression, leading to the trough in Fig.~\ref{fig:trough_DE}.

\begin{figure}
      \includegraphics[width=0.48\textwidth]{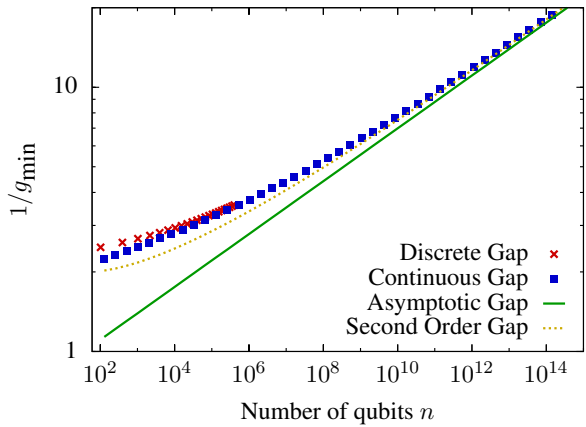}
      \caption{
Spectral gap            $\gmin$ vs.\ $n$ for $\alpha=3/10$ for our three different gap extraction methods with a square barrier.  This plot covers a much larger range than Fig.~\ref{fig:comparison}, allowing us to see that the continuous gap does eventually approach the asymptotic gap.  However, this approach occurs at extremely large $n$ that are impractical for most current applications.  In addition, we have included the second order expansion of the asymptotic large $n$ expression from Eq.~\ref{eq:next_order} that more closely approximates the continuous gap.
      }
      \label{fig:comparison_big}
\end{figure}

With our current computational limits, we cannot push exact diagonalization of the symmeterized $n$ qubit Hamiltonian to much higher $n$, but the condition in Eq.~\ref{eq:trans_con} can be solved for extremely large $n$ with rounding error being the only block.  Therefore, we can extend Fig.~\ref{fig:comparison} to higher $n$ to get Fig.~\ref{fig:comparison_big} from which we can see that the continuous gap does eventually approach its asymptotic value, but only for extremely large $n> 10^{12}$.

Fig.~\ref{fig:comparison_big} also shows another curve, labeled ``Second Order Gap.''  This curve shows the asymptotic expression for the gap in the large-$n$ limit, but it keeps both the leading order term and the second highest order term so that this second order gap expression is given by
\begin{equation}
      \label{eq:next_order}
      \gmin^{(2^{\text{nd}}\text{O})} = \frac{8\omega^{1/2}}{c\sqrt{\pi}}\epsilon^{2\alpha-1/2} - \frac{16}{c\pi}(\ln 2+\psi_0(-\frac{1}{2}))\epsilon^{4\alpha-1}
\end{equation}
for $1/4<\alpha<1/3$ where $\psi_m(z)$ is the polygamma function.

The second order term does improve agreement with the continuous gap, but this fact also shows that the second order (and higher) terms are still relevant and important for determining run time behavior even at the very large end of what quantum computers are expected to be able to accomplish in the next few decades.  Since even a simple tunneling problem needs more than just its asymptotic limit to accurately describe its behavior at any reasonable $n$, this data casts doubt on the efficacy of asymptotic analysis.

We have exclusively used parabolic wells up to this point so the argument could be made that this large-$n$ behavior is an artifact of our choice of wells.  To counter this argument, we have done this same analysis using an appropriately sized square well instead of a parabolic well.  The asymptotic scaling behavior of $\gmin$ has the same $n$ and $\alpha$ dependence in the square well case, and extremely large $n$ are still needed for the asymptotics to accurately describe the exact gap for the square well.

As a general rule, we have noted that the smoother the potential is the faster $\gmin$ begins to approximate its asymptotic value.  We have chosen the quadratic potential since it makes the differential equation solvable, but we have run a few other simulations with different shaped barriers.  While a smoother barrier (such as a Gaussian) does level off faster, it still takes significantly large $n$ to have approximately the correct slope.  Exact analysis is more difficult for different barrier and potential shapes since we cannot get exact asymptotic scaling expressions for them up to the constant prefactors.

\subsection{Ratio of Exact Gap to Asymptotic}

\begin{figure}
      \includegraphics[width=0.48\textwidth]{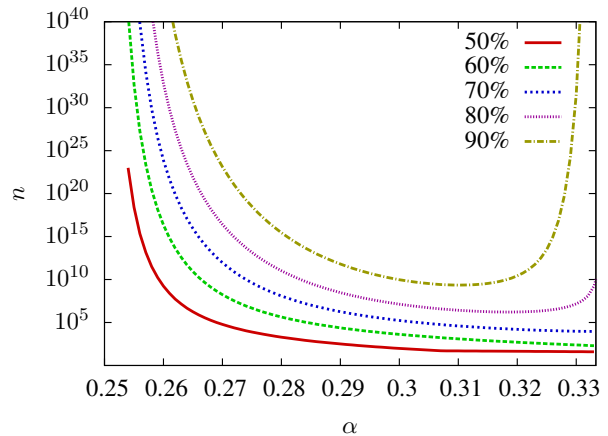}
      \caption{
            We show the $n$ necessary for the ratio of the continuous gap and the asymptotic gap to reach the indicated percentages as a function of $\alpha$.  Especially close to the boundaries of the polynomial region, the system requires extremely large $n$ to reach its asymptotic value.  When coupled with Fig.~\ref{fig:trough_DE}, this plot indicates that for the middle values of $\alpha$ the continuous gap reachs its asymptotic value faster by deviating more from its asymptotic scaling.
      }
      \label{fig:ntoreach}
\end{figure}

We may also want to know what size of $n$ is necessary before the continuous gap is a certain fraction of the asymptotic gap.  In Fig.~\ref{fig:ntoreach}, we plot the $n$ necessary for the ratio of the continuous gap to the asymptotic gap to reach the specified value.  When we want 50\%, the asymptotic gap is a good approximation relatively quickly for higher $\alpha$, but even for lower $\alpha$, extremely large $n$ are required.  For ratios around 80\% or 90\% the necessary $n$ becomes prohibitively large, even for the best $\alpha$ values.  Additionally, for these higher ratios, high $\alpha$ also start requiring astronomical $n$ in addition to the low $\alpha$ values.

The behavior noted in Fig.~\ref{fig:ntoreach} at first seems to contradict the information in Fig.~\ref{fig:trough_DE} since these two plots claim that opposite regions are worse.  This is not a contradiction because the bad cases in Fig.~\ref{fig:trough_DE} have scaling behavior very different from the asymptotic behavior that paradoxically allows them to approach the asymptotic value of the gap faster.  Thus, the middle range of $\alpha$ approachs the true asymptotic values of the gap faster at the cost of having vastly different scaling behavior until they reach the appropriate value.  However, this faster approach to the asymptotic value is still extremely slow in comparison to the concerns of practical quantum computers, meaning that the benefit of a faster approach cannot be realized.

To explore this phenomena more, let us analytically consider the ratio between the leading order asymptotic gap and the second order asymptotic gap
\begin{align}
      \frac{\gmin^{(2^{\text{nd}}\text{O})}}{\gmin^{(1^{\text{st}}\text{O})}} &= 1 - \frac{2}{\sqrt{\pi\omega}}(\ln 2+\psi_0(-\frac{1}{2}))\epsilon^{2\alpha-1/2} \nonumber\\&= 1 - \kappa \epsilon^{2\alpha-1/2},
\end{align}
where $\kappa \coloneqq \frac{2}{\sqrt{\pi\omega}}(\ln 2+\psi_0(-\frac{1}{2})) \approx 0.8233/\sqrt{\omega}$.

In our case $\omega = 4/3$, so the prefactor, $\kappa$, is close to $0.7$, meaning any galactic issues with this expression are a direct result of the $\epsilon$ dependence.  If we want this ratio to reach some threshhold value $0<v<1$, then the necessary $n(v)$ becomes
\begin{equation}
      \label{eq:nofv}
      n(v) = 2\left(\frac{\kappa}{1-v}\right)^{1/(2\alpha-1/2)}.
\end{equation}
Since $\kappa\approx 0.713$, the expression $\frac{\kappa}{1-v}$ is greater than one for most possible values of $0<v<1$, so the size of $n(v)$ depends mostly on the exponent.  When $2\alpha-1/2$ is close to zero, meaning the problem is close to the boundary between the constant and polynomial scaling regions, the exponent is extremely large, leading to large $n$ necessary to reach a specific $v$.

However, the expression in Eq.~\ref{eq:nofv} would seem to indicate that the necessary $n$ should be relatively small for $2\alpha-1/2$ farther from zero as we are close to the boundary between the polynomial and exponential scaling regions.  Therefore the second order expansion cannot explain the large spike in $n(v)$ seen for high $2\alpha-1/2$ in Fig.~\ref{fig:ntoreach} for large $v$.

\section{Polynomial vs.\ Exponential Scaling Issues}
\label{sec:scaling}

Given the discrepancies between the actual gap and the asymptotic gap, an important question becomes how easily scaling behavior can be mislabeled.  Given a data set, coming from an experiment or numerical simulation, how difficult is it to tell apart a polynomial scaling algorithm from an exponential or super-polynomial algorithm.  Fig.~\ref{fig:trough_DE} already explored how easily the asymptotic scaling power could be misidentified, but that analysis assumed a power law fit from the beginning.  In this section, we explore ideas for analyzing how much a data set follows a power law model or an exponential model to judge if finite data indicates polynomial or super-polynomial scaling with $n$.

Whereas Fig.~\ref{fig:trough_DE} looked at a broad sweep of data in an attempt to extract scaling information, this section will be concerned with small scale features of curves, asking how well a curve can locally be approximated by a certain functional form.  Our ultimate goal is to be able to take data like that presented in Fig.~\ref{fig:comparison} and judge how likely it is that the data is obeying a power law or exponential relationship.

We will look at both a power law and exponential relationship, and for ease of writing we define $f\coloneqq \ln\gmin$ and $x\coloneqq \ln n$:
\begin{align}
      \gmin = A n^{-p}     &     \Rightarrow            f_{\text{poly}} = \ln A - p x,\\
      \gmin = B \exp(-C n^q)  &     \Rightarrow            f_{\text{exp}} = \ln B - C e^{q x}.
\end{align}
Our analysis technique relies on properties of $f$ and its derivatives with repspect to $x$.  In Fig.~\ref{fig:trough_DE}, when we did linear fits to log-log data, we were already looking at the properties of $f'(x)$.

Two natural other quantities to look at would be $f''(x)$ and the curvature of the function $f(x)$.  The curvature is not a good metric since the curvatures of both $f_{\text{poly}}$ and $f_{\text{exp}}$ go to zero for $x\to\infty$.  The second derivative $f''(x)$ is a better metric  since $f_{\text{poly}}''(x)$ goes is zero; whereas, $f''_{\text{exp}}(x)$ goes to $-\infty$ as $x\to\infty$.  The problem with $f''(x)$ is that we cannot immediately extract any additional information from it about the exponential.

\subsection{Quality of Fits}

Our proposed analysis technique relies on another important feature of the derivatives of these log-log functions.  Specifically, we can look at the ratio, $R$, of the second derivative to the derivative:
\begin{align}
      R_{\text{poly}} &= \frac{f''_{\text{poly}}(x)}{f'_{\text{poly}}(x)} = 0\nonumber\\ R_{\text{exp}}&=\frac{f''_{\text{exp}}(x)}{f'_{\text{exp}}(x)} = q.
\end{align}
Note that only exponential scaling gives a non-zero, constant derivative ratio which gives us another metric to judge if a function has exponential scaling and allows us to discover $q$.  In general, sub-exponential scaling takes this ratio to zero for large $n$.  True polynomial scaling has a ratio of exactly zero, but given that we do not immediately reach asymptotic scaling, asymptotic polynomial and sub-exponential scaling are not immediately distinguishable through this method.

\begin{figure}
      \begin{center}
            \includegraphics[width=0.48\textwidth]{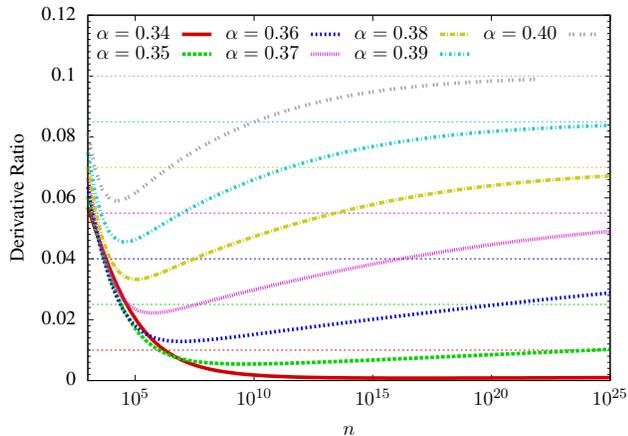}
      \end{center}
      \caption{
            The derivative ratio $R = f''(x)/f'(x)$ where $f\coloneqq \ln\gmin$ and $x\coloneqq \ln n$ for the continuous gap.  For true exponential scaling, this ratio should approach a constant equal to the power of $n$ in the exponential.  The dashed lines are what this ratio should be asymptotically.  None of these curves do a good job approximating their asymptotic constant value, and even the very large barrier cases require large $n$ to even approach the correct constant.  This plot indicates that even the exponential scaling region does not approach its asymptotic expression at reasonable $n$.
      }
      \label{fig:exp_ratio}
\end{figure}

In Fig.~\ref{fig:exp_ratio}, we have plotted the ratio $R = f''(x)/f'(x)$ for the continuous gap for $\alpha$ values that should have asymptotically exponential scaling.  The thin dashed lines represent what the curve of the same color should approach in its asymptotic limit.  For lower $\alpha$, close to the boundary with the polynomial region, the $R$ vs.\ $n$ curves do not come close to their asymptotic values, even for the extremely large $n$ shown, up to $n=10^{25}$.  Some of the more strongly exponential systems do approach their asymptotic values for the $n$ displayed, but these systems still require extremely large $n$.

Based on the nature of the ratio $R$ as a function of $x= \ln n$, an asymptotically increasing or non-zero constant ratio would imply a super-exponential or exponential scaling of $\gmin$ with $n$, so it could be argued that these plots imply non-polynomial behavior much earlier.  Even with this argument, many of the borderline $\alpha$ cases continue decreasing for a large range of $n$ and do not have an upturn until much larger $n$.  Additionally, if our goal is to extract information on the type of exponential scaling, Fig.~\ref{fig:exp_ratio} implies that we need very large $n$.

\begin{figure}
      \begin{center}
            \includegraphics[width=0.48\textwidth]{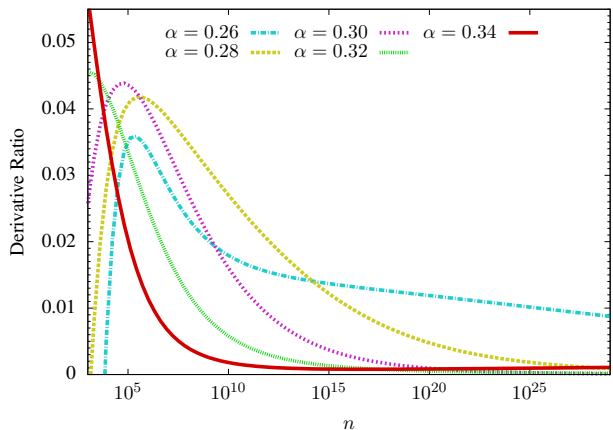}
      \end{center}
      \caption{
            The derivative ratio $R = f''(x)/f'(x)$ where $f\coloneqq \ln\gmin$ and $x\coloneqq \ln n$ for the continuous gap.  These $\alpha$ values were chosen in the polynomial scaling region, and for true power law scaling, we would expect the ratio $R=0$.  We do see the curves approaching zero, but all of them require extremely large $n$ to do so, with several of the smaller barrier cases being far away from zero even at the extreme end of our simulation range.  The solid, red line is for $\alpha=0.34$ in the exponential region; notice that the ratio appears to approach zero faster than many of the ratios for $\alpha$s in the polynomial scaling region.
      }
      \label{fig:poly_ratio}
\end{figure}

Asymptotically polynomial scaling should give us an $R$ ratio that is zero, so in Fig.~\ref{fig:poly_ratio}, we have plotted $R$ as a function of $n$ for several $\alpha$ values that asymptotically have polynomial scaling.  All of these curves, except the solid line for $\alpha=0.34$, should approach zero, but as we see, many of them take extremely large $n$ to actually approach this asymptotic value.  Notice especially that $\alpha$ closer to $1/4$, the boundary between constant and polynomial scaling, require even larger $n$ to reach zero.  This behavior of small $\alpha$ corresponds well with what is seen in Fig.~\ref{fig:ntoreach} in that lower $\alpha$ take a long time to reach the correct asymptotic value.

Also notice that the solid, red line for $\alpha=0.34$ in Fig.~\ref{fig:poly_ratio} appears to approach zero faster than many of the ratios for $\alpha$ actually in the polynomial scaling region.  Therefore, for a large range of $n$, it would be easier to interpret $\alpha=0.34$ as having a power law scaling behavior than many of the $\alpha$ close to $1/4$ which take very large $n$ to display the appropriate ratio behavior.

The slowness of lower $\alpha$ systems in reaching their asymptotic value is at least partly due to the size of their scaling powers.  Fig.~\ref{fig:trough_DE} indicates that these $\alpha$ values close to the constant boundary reach their asymptotic scaling behavior more quickly.  However, their asymptotic scaling behavior is a very small power law, meaning that any change in these functions will be small.  By reaching their asymptotic scaling earlier, it will take these low $\alpha$ systems even longer to reach the correct asymptotic value for their scaling, as seen in Figs.~\ref{fig:ntoreach} and \ref{fig:poly_ratio}.

\section{Conclusion}
\label{sec:conc}

The problem of tunneling through a barrier in a quadratic well is a simple quantum mechanical system with elements that are not unfamiliar to an introductory quantum mechanics class, so it is surprising that asymptotics have difficulty describing such an elementary system at reasonable system sizes.  Asymptotic analysis is deemed useful because it is assumed that the run time of an algorithm  is well approximated by its asymptotic behavior starting at relatively small system sizes, $n$.

Our study shows that barrier tunneling is a case where asymptotic behavior is not a good descriptor of spectral gap and therefore run time scaling until extremely large $n$.  This failure of asymptotic analysis is worsened by the direction of the failure: asymptotically polynomial algorithms appear to locally scale in a super-polynomial manner as seen in Fig.~\ref{fig:poly_ratio}, making it more difficult to justify the use of asymptotic analysis for small system sizes.

Quantum tunneling is a key part of many near-term quantum algorithms, including quantum adiabatic optimization.  These algorithms are gaining popularity because of their applicability on smaller, possibly non-error-corrected, quantum computers.  Our results on the difficulty of describing barrier tunneling in small systems with asymptotic analysis suggest that care should be taken when analyzing the potential power of these algorithms for near-term experiments and systems.

\subsubsection*{Acknowledgements}
This material is based upon work supported by the National Science Foundation under Grants No.\ 1314969
and No.\ 1620843.



\begin{thebibliography}{99}


\bibitem{Farhi2000} E. Farhi, J. Goldstone, S. Gutmann, M. Sipser, quant-ph/0001106 (2000).

\bibitem{Farhi2002} E. Farhi, J. Goldstone, S. Gutmann, quant-ph/0201031 (2002).

\bibitem{Farhi2008} E. Farhi, J. Goldstone, S. Gutmann, D. Nagaj. Int. J. Quantum Inf. {\bf 6}, 3 (2008).

\bibitem{Martonak} R. Marto\u{n}\'{a}k, G. E. Santoro, E. Tosatti. Phys. Rev. B {\bf66}, 094203 (2002).

\bibitem{Hastings} M. B. Hastings, M. H. Freedman, Quant. Inf. \& Comp. {\bf 13}, 11-12 (2013).

\bibitem{Boxio2} S. Boixo, T. F. R{\o}nnow, S. V. Isakov, Z. Wang, D. Wecker, D. A. Lidar, J. M. Martinis, M. Troyer, Nature Phys. {\bf10}, 218 (2014).

\bibitem{Battaglia} D. Battaglia, G. Santoro, E. Tosatti, Phys. Rev. E {\bf71}, 066707 (2005).

\bibitem{Crosson} E. Crosson, M. Deng, quant-ph/1410.8484 (2014).

\bibitem{Harrow} E. Crosson, A. Harrow, quant-ph/1601.03030  (2016).

\bibitem{Heim} B. Heim, T. F. R{\o}nnow, S. V. Isakov, M. Troyer, Science {\bf 348}, 6231 (2015).

\bibitem{Muthukrishnan} S. Muthukrishnan, T. Albash, D. A. Lidar, Phys. Rev. X {\bf 6}, 031010 (2016).

\bibitem{Brady} L. Brady, W. van Dam, Phys. Rev. A {\bf 93}, 032304 (2016).

\bibitem{Kong} L. Kong, E. Crosson, quant-ph/1511.06991 (2015).

\bibitem{Jiang}  Z. Jiang, V. N. Smelyanskiy, S. V. Isakov, S. Boixo, G. Mazzola, M. Troyer, and H. Neven, quant-ph/1603.01293 (2016).
(2016).

\bibitem{Aharonov} D. Aharonov, W. van Dam, J. Kempe, Z. Landau, S. Llyod, O. Regev, SIAM J. Comp. {\bf 37}, 166 (2007).

\bibitem{Brady2} L. Brady, W. van Dam, Phys. Rev. A {\bf 94}, 032309 (2016).

\bibitem{Boxio3} S. Boxio, S. V. Isakov, V. N. Smelyanskiy, R. Babbush, N. Ding, Z. Jiang, J. M. Martinis, H. Neven, quant-ph/1608.00263 (2016).

\bibitem{Jansen} S. Jansen, M. Ruskai, R. Seiler, J. Math. Phys. {\bf48}, 102111 (2007).

\bibitem{Goldstone} J. Goldstone (unpublished).

\bibitem{Villain} J. Villain, J . Phys. France, {\bf35}, 27 (1974).

\bibitem{Reichardt} B. W. Reichardt, in Proceedings of the 36th Annual ACM Symposium on Theory of Computing (STOC'04), ACM Press(2004).

\end{thebibliography}
\end{document}